\documentstyle[12pt]{article}

\def\be{\nopagebreak[3]\begin{equation}}
\newcommand{\ee}{\end{equation}}
\def\ba{\begin{array}}
\def\ea{\end{array}}
\def\dmas{\partial_{+}}
\def\dmenos{\partial_{-}}
\def\dpm{\partial_{\pm}}

\def\bea{\begin{eqnarray}}
\def\eea{\end{eqnarray}}
\def\caption#1{\vskip 0.1in\centerline{\vbox{\hsize 2in\noindent
     \tenpoint\baselineskip=14pt\strut #1\strut}}}

\begin{document}

\begin{titlepage}
\begin{flushright}
CERN-TH-7204/94\\
March, 1994\\
hep-th/9403155
\end{flushright}
%\vspace*{20pt}

\bigskip

\begin{center}
{\LARGE On Non-Abelian Duality}

\vskip 0.7truecm

{Enrique \'Alvarez
\footnote{On leave of absence from: Departamento de F\'{\i}sica
Te\'orica,
Universidad Aut\'onoma de Madrid, 28049 Madrid, Spain},
Luis \'Alvarez-Gaum\'e
and  Yolanda Lozano \footnotemark[1]

\vspace{1pc}

{\em Theory Division CERN \\1211 Geneva 23\\ Switzerland}}\\

\vspace{1pc}

{\large \bf Abstract}
\end{center}

A general study of non-abelian duality is presented. We first
identify a possible obstruction to the conformal invariance
of the dual theory for non-semisimple groups. We construct
the exact non-abelian dual for any Wess-Zumino-Witten (WZW)
model for any anomaly free subgroup, and the corresponding
extension for coset conformal field theories. We characterize
the exact non-abelian dual for $\sigma$-models with chiral
isometries and extend the standard notion of duality to
anomalous subgroups of WZW-models, thus giving a way of
constructing dual transformations for different groups on
the left and on the right. We also present some new examples
of non-abelian duality for four-dimensional gravitational
instantons.

\vfill
\end{titlepage}

\def\theequation{\thesection . \arabic{equation}}

\section{Introduction}
\setcounter{equation}{0}

Duality transformations with respect to abelian
(discrete or continuous) symmetries
have a long history in statistical mechanics \cite{r1}.
In String Theory and two-dimensional
Conformal Field Theory abelian duality with respect to
continuous symmetries has received an increasing amount of
attention \cite{r2}.
Of more recent history is the notion of non-abelian
duality \cite{r3,r4}, which has no analogue in
statistical mechanics. The basic idea of \cite{r3},
partly inspired in the treatment of abelian duality
presented in \cite{r5}, is to consider a conformal
field theory with a non-abelian symmetry group $G$.
Particularly one considers a $\sigma$-model on a
manifold $M$ with isometry group $G$. The duality
transformation proceeds in two steps: i) First one
gauges the isometry group, thus introducing some
gauge field variables $A_{\mu}^a$. The gauge field
is required to be flat; and this is implemented by
adding to the lagrangian a term of the form
$\chi_a F_{\mu \nu}^a \epsilon^{\mu \nu}$,
where $\chi_a$ is a lagrange multiplier imposing
the flatness constraint on the gauge connection.
It is naively clear that if we first perform the
integral ovel $\chi_a$, this provides a $\delta$-function
$\delta(F^a)$ on the measure, which in turn implies that
$A_{\mu} = g^{-1}\partial_{\mu}
g$ is a pure gauge (we consider a
spherical world-sheet for simplicity). Using the original
symmetry of the theory we can absorb the gauge field by a
 change of variables on the manifold. In this way we recover
the original model. ii) The second step consists of
integrating first the gauge field $A_{\mu}^a$.
Since we have no gauge kinetic term, the integration
is gaussian, and we obtain a lagrangian,
$\tilde{L}(\phi,\chi)$, depending on the
original variables $\phi^i$ and the auxiliary
variables $\chi_a$, which is still gauge invariant.
Fixing finally the gauge we obtain the dual theory.
In the abelian case it is also possible to work out
the mapping between operators for the original and
the dual theory, as well as the global topology of
the dual manifold \cite{r2,r4}. In \cite{r5} it was
further shown that if one starts with a conformal
field theory (CFT), conformal invariance is preserved
by abelian duality. The proof was based on an analogy
between the duality transformation
and the GKO construction \cite{r6}. Thus for $G$
abelian we have a rather thorough understanding of
the detailed local and global properties of duality.
Up till now
our knowledge of the non-abelian case was far more
limited, and it is one of the purposes of this paper
to mend this situation.

An interesting example of non-abelian
duality recently studied is the one presented
in \cite{r7}. They considered a cosmological
solution to string theory of Bianchi type V.
In fact to satisfy the $\beta$-function constraints
for conformal invariance, the space is flat.
It is the interior of the light-cone through the origin in
Minkowski space. They then performed the non-abelian
duality transformation with respect to a non-semisimple
subgroup of the Lorentz group, and to their
surprise they found that the dual model did not
satisfy conformal invariance. We will show that
when one analyzes carefully the measure of integration
over the gauge fields and its dependence on the world
sheet metric, one encounters a mixed gauge and
gravitational anomaly \cite{r8} when any generator
of $G$ in the adjoint representation has a non-vanishing
trace\footnote{The breakdown of conformal invariance for
non-semisimple groups due to the possible existence of
generators in the adjoint representation which are not
trace free was anticipated in \cite{givroc} in the note added.}.
This only happens for non-semisimple groups.
This mixed anomaly generates a contribution to
the trace anomaly which cannot
be absorbed in a dilaton shift. This explains
why the dual model in \cite{r7} violates conformal
invariance, and imposes a mild anomaly cancellation
condition for the consistency of non-abelian duality.

The organization of this paper is as follows:
In section two we briefly summarize the operations
involved in explicitly computing the abelian or
non-abelian dual of a given theory. We will show
that one can understand the transformations as a
change of gauge condition, and this will make clear
that one of the possible obstructions to conformal
invariance is found in mixed two-dimensional anomalies.
In section three we compute the dependence of the gauge
field measure on the world-sheet metric. We evaluate
the leading terms in the effective action, and we
explicitly show the existence of mixed anomalies
when some generators of the adjoint representation
of the isometry group are not trace-free.
Following the approach developed in this section,
we present in section four an exact treatment of the
non-abelian duality transformation for Wess-Zumino-Witten
(WZW) models \cite{r9} with group $G$ with respect to
any vector subgroup of $G \times G$. If $k$ is the
level of the WZW model, the non-abelian dual with respect
to  $H \subset G_D$
(where $G_D$ is the diagonal embedding of $G$)
 is given by $(G/H)_k \times H_k$
where $(G/H)_k$ is the coset GKO construction for
$G$ and $H$ \cite{r6} and $H_k$ is a WZW theory
for $H$ with level $k$ (assuming for simplicity
that the Dynkin index for the embedding
$H \subset G$ is $1$). In this way we also
find the global topology of the non-abelian dual.
The extension of our construction to the coset conformal
field theory $G/H$ when $H$ has a non-trivial centralizer
in $G$ is straightforward. In section five we
consider a general $\sigma$-model with left and right
chiral non-abelian currents, and show that the
non-abelian dual with respect to any vector subgroup
is again described by a generalized GKO construction
coupled to a WZW theory determined by the auxiliary field
$\chi$.  This will give us the clue of how to generalize
duality in  a heterotic fashion even with respect to
anomalous isometry groups.  The auxiliary field
is given transformation properties to cancel
the anomaly as in the abelian case.
 We then make some remarks on the general case
where the isometry currents are not necessarily chiral.
Once again the key to the exact treatment of
non-abelian duality lies on the coupling to a WZW theory.
One of the basic ingredients is a way of writing the
term $\chi_a F_{\mu \nu}^a \epsilon^{\mu \nu}$
(which always appears in duality transformations)
in terms of WZW lagrangians. In section six we
present some interesting non-abelian duality
transformations for the Eguchi-Hanson and
Euclidean Taub-NUT gravitational instantons.
We finish with the conclusions and outlook in section seven.

\section{Duality transformations and gauge equivalence}
\setcounter{equation}{0}

We work throughout with light-cone variables
in two dimensions, $x^{\pm} = \tau \pm \sigma$.
Consider first a $\sigma$-model with an isometry
generated by the Killing vector $\xi^i$
\be
L_0 = {1\over 2\pi}g_{ij}(\phi) \partial_{+}\phi^i
\partial_{-}\phi^j
\ee
and
\be
\pounds(\xi) g_{ij} = \nabla_i \xi_j + \nabla_j
\xi_i = 0
\ee
Gauging the isometry and adding a Lagrange multiplier
$\chi$ to force the gauge connection to be flat we
obtain the starting point of any  duality transformation:
\be
L = {1\over 2\pi} g_{ij}(\phi) D_{+} \phi^i D_{-}\phi^j +
 {1\over 2\pi} \chi (\partial_{+} A_{-} - \partial_{-} A_{+})
\ee
\be
D_{\pm} \phi^i = \partial_{\pm} \phi^i + \xi^i(\phi) A_{\pm}
\ee
If we integrate first over $\chi$, we obtain (on the sphere)
$A_{\pm} = \partial_{\pm} \alpha$. Then we make a change
of variables on $\phi^i$ to eliminate $\alpha$. This is
done as follows: The $\sigma$-model is defined on a
manifold $(M,g)$, the metric being constant along
the orbits of the isometry group, $G$. The action of
$G$ on $M$ is represented by a function $\phi^{'}\,^i =
f^i (\phi,\alpha)$. Changing variables from $\phi^i$
to $\phi{'}\,^i$ brings the action into the original
form with respect to the $\phi^{'}$ variables. Since
the path integral measure contains a factor of
$(det g)^{1\over 2}$ at each point, i.e.,
$\prod_{\sigma,\tau} {\cal D}\phi(\tau,\sigma)
(detg(\phi(\tau,\sigma)))^{1\over 2}$  one indeed
recovers the same quantum theory we started with.
If we now integrate over the gauge fields, we obtain
the dual action. Choosing coordinates adapted to
$\xi^i$, $\phi^i =(\theta,\phi^{\alpha})$, $\xi^i
{\partial \over \partial \phi^i} =
{\partial \over \partial \theta}$, the dual lagrangian
is \cite{r2}:
\be
\tilde{L} = {1\over 2\pi} (\tilde{g}_{ij}(\tilde{\phi}) +
\tilde{b}_{ij}(\tilde{\phi}))\partial_{+} \tilde{\phi}^i
\partial_{-} \tilde{\phi}^j
\ee
with $\tilde{\phi}^0 = \chi$, $\tilde{\phi}^{\alpha} =
\phi^{\alpha}$, and
\begin{eqnarray}
\tilde{g}_{00}& =& {1\over g_{00}}\nonumber\\
\tilde{g}_{\alpha \beta}& = &g_{\alpha \beta}-
{g_{0\alpha}g_{0\beta} \over g_{00}}\nonumber\\
\tilde{b}_{0\alpha}& =& {1\over g_{00}}g_{0\alpha}
\nonumber\\
\tilde{b}_{\alpha \beta}&=& 0.
\end{eqnarray}
We can look at this transformation differently.
Since the lagrangian is invariant under
\begin{eqnarray}
\delta\phi^i& =& \epsilon \xi^i(\phi)\nonumber\\
\delta A_{\pm}&=& - \partial_{\pm}\epsilon
\end{eqnarray}
we can choose the light-cone gauge, $A_{+} = 0$,
yielding
\be
L = L_0 + {1\over 2\pi} \xi_i(\phi)\partial_{+}
\phi^i A_{-} + \chi \partial_{+} A_{-}
\ee
ignoring the ghost contribution.
The integration over $A_{-}$ gives a $\delta$-function
imposing $\partial_{+}\chi -\xi_i \partial_{+}
\phi^i = 0$. The integration over $\chi$ is now trivial.
Hence our lagrangian in the $A_{+} =0$ gauge is
equivalent to the original model.
If we now choose instead the Landau gauge
$\partial_{\mu}A^{\mu} =0$, we can parametrize
(on the sphere) $A_{\mu} = \epsilon_{\mu\nu}\partial^{\nu}
\rho$; that is, $A_{\pm} = \pm\partial_{\pm}\rho$.
The lagrangian now reads in adapted coordinates:
\bea
L &=& g_{00}(\dmas \theta + \dmas \rho)(\dmenos
\theta - \dmenos \rho) \nonumber\\
&+& g_{0\alpha} (\dmas \theta + \dmas \rho)\dmenos
\phi^{\alpha} \nonumber\\
&+& g_{0\alpha}( \dmenos\theta - \dmenos \rho)
\dmas\phi^{\alpha} \nonumber\\
&+& g_{\alpha \beta} \dmas \phi^{\alpha}
\dmenos\phi^{\beta} + \dmas\chi\dmenos\rho +
\dmenos\chi\dmas\rho
\eea
After some simple manipulations we obtain:
\bea
2\pi L &=& g_{00}(\dmas \theta + \dmas \rho +
{1\over g_{00}}(j_{+} - \dmas\chi))
(\dmenos \theta - \dmenos \rho +{1\over g_{00}}(j_{-} +
\dmenos\chi))\nonumber\\
&-&(\dmas\theta\dmenos\chi - \dmenos\theta\dmas\chi) +
{1\over g_{00}} \dmas\chi\dmenos\chi\nonumber\\
&+&(g_{\alpha\beta} - {g_{0\alpha} g_{0\beta}\over
g_{00}})\dmas\phi^{\alpha}\dmenos\phi^{\beta}\nonumber\\
&-&{g_{0\alpha}\over g_{00}} (\dmas\phi^{\alpha}
\dmenos\chi - \dmenos\phi^{\alpha}\dmas\chi)
\eea
where $j_{\pm}=g_{0\alpha}\partial_{\pm}\phi^{\alpha}$.
Once again we have ignored the Faddeev-Popov ghosts.
The second term in the preceding equation is purely
topological and it sets the global range of the auxiliary
field $\chi$. The last three terms give the dual lagrangian,
whereas the first term can be eliminated by integrating
over $(\theta,\rho)$, yielding a determinant of
$1/g_{00}$, which determines the duality properties
of the dilaton, if one defines it in a convenient
way, as done by Buscher [2].

If the theory we started with is gauge invariant
at the quantum level, then a change of gauge is not
a change in physics, and both models are equivalent.
This argument has the advantage of holding even for
theories which are not conformally invariant. It is
also clear that the same construction applies to
non-abelian duality. Within this point of view,
possible problems with duality in CFT should be
found in the interplay between the trace or the
gravitational anomaly with the gauge anomaly,
and this will be our main theme in section three.

For future reference we collect here all the basic
formulae for non-abelian duality. This reduces to
the construction of a gauge $\sigma$-model with
arbitrary WZW-term, where we gauge the isometry
group $G$. This has been carried out in \cite{r11,r12}.
The general form of a WZW $\sigma$-model on a manifold
$M$ with metric $g_{ij}$ is:
\be
S[\phi] = {1\over4\pi} \int_{M_2} dx^{+}dx^{-}
g_{ij}(\phi)\dmas\phi^i\dmenos\phi^j + {1\over 12\pi}
\int_{B_3,\partial B_3 =M_2} H
\ee
H is a three-form on $M$ which belongs to the integral
cohomology of $M$, and we assume it normalized so that
\be
{1\over 24\pi^2} \int_{C_3} H = {\rm integer}
\ee
where $C_3$ is a 3-cycle in $M$. $B_3$ is a
three-dimensional space whose boundary is $M_2$.
If $M_2$ has a trivial topology, $B_3$ is topologically
 a three-ball. However, if $M_2$ is a Riemann surface
of genus g (so that $x^{\pm}$ are complex coordinates),
then $B_3$ is a genus g handlebody.
The measure $dx^{-}\wedge dx^{+} = 2 d\tau\wedge d\sigma$
is chosen for convenience. The normalization $1/4\pi$
implies that the two-point function for the massless
scalar field is $ - log(x^{+} - y^{+})(x^{-} - y^{-})$.
The isometry group $G$ of $M$ is generated by the
Killing vectors $\xi_{a}\,^i$
($a=1\dots {\rm dim\, G}$).
Introducing gauge fields $A_{\pm}\,^a$, the gauge
transformations are:
\bea
&&\delta\phi^i = \epsilon^a(x) \xi_{a}\,^i(\phi)\nonumber\\
&&\delta A_{\pm}^a = \dpm \epsilon^a + f^a\,_{bc}A_{\pm}\,^b
\epsilon^c\nonumber\\
&&D_{\pm} \phi^i = \dpm \phi^i - \xi_{a}\,^i A_{\pm}\,^a
\eea
and $f^a\,_{bc}$ are the structure constants of the group
\be
[\xi_a,\xi_b]^i = (\xi_a\,^k {\partial\xi_b\,^i
\over\partial\phi^k} - \xi_b\,^k{\partial\xi_a\,^i
\over\partial\phi^k}) = f _{ab}\,^c \xi_c\,^i
\ee
Global invariance of the action implies some
conditions on $g_{ij}$ and $H_{ijk}$. On the metric
the conditions are just the Killing equations. The
invariance of the torsion is, in turn, conveyed by
the equations
$$
{\cal L}(\xi_{a}) H = 0
$$
Since $H$ is closed, and acting on forms ${\cal L}
(\xi_{a}) = i(\xi_{a})d + di(\xi_{a})$ ($i(\xi_a)$
being the interior product, or contraction with the
vector field $\xi_a$), the invariance now requires
that for every $\xi_a$ we should be able to find a
one-form globally defined on $M$ and satisfying
$i(\xi_a) H = dv_a$. In components:
\be
\xi_a\,^i H_{ijk} = \partial_j v_{ak} -
\partial_k v_{aj}
\ee
We need furthermore the condition that the $v_{ai}$
form a representation of the isometry group
\be
{\cal L}(\xi_{a}) v_{bi} = \xi_{a}\,^{k}\partial_{k}
v_{bi}+v_{bk}\partial_{i}
\xi_{a}\,^{k} = f_{ab}^{c} v_{ci}
\ee
The simplest way to gauge the preceding action is
to use Noether's procedure. We first perform a
variation of the ungauged action. Since the
parameters are point-dependent, there will be
terms in it proportional to $\dmas \epsilon^a$
and $\dmenos\epsilon^a$:
\be
4\pi \delta S[\phi] = \int \dmas\epsilon^a(\xi_{ai}+
v_{ai})\dmenos\phi^i + \dmenos\epsilon^a(\xi_{ai} -
v_{ai})\dmas\phi^i
\ee
Note in passing that the condition to have a left-chiral
current ($\dmenos J_{+a} = 0$) is $\xi_a + v_a =0$;
similarly a right chiral current requires $\xi_a-v_a=0$.
To cancel our variation, we add a term to the action linear
in the gauge fields $A_{\pm}\,^a$. In turn this new
contribution to the action will have a gauge variation,
which we try to cancel by adding
another term to the action which is quadratic in $A_{\pm}$.
At this point one finds that the variation of the full action
is independent of the $\phi$-fields and it is equal to the two
dimensional gauge anomaly. The gauged action \cite{r11,r12}
becomes:
\bea
4\pi S[\phi,A]&=&\int_{M_2}d x^-d x^+ g_{ij}(\phi)D_{+}\phi^i D_{-}
\phi^{j}+\frac13 \int_{B_3} H_{ijk}d\phi^i
d\phi^j d\phi^k \nonumber \\
&-&\int (A_+^a v_{ai}\partial_-\phi^i-A_
-^a v_{ai}\partial_+\phi^i)+\int c_{[ab]} A_+^a A_-^b,
\eea
where
\be
c_{ab}=v_{ai}\xi_b^i, \,\, c_{(ab)}=\frac12 (c_{ab}+c_{ba}),
\,\, c_{[ab]}=\frac12 (c_{ab}-c_{ba}),
\ee
and under a gauge variation:
\be
\delta S[\phi,A]=-\frac{1}{4\pi}\int c_{(ab)}(A_+^a\partial_-
\epsilon^b
-A_-^a\partial_+\epsilon^b),
\ee
The coefficients $c_{(ab)}$ are constant:
$$
dc_{ab}=d(v_a.\xi_b)=d(i(\xi_b).v_a)={\cal L}(\xi_b)
.v_a-i(\xi_b)i(\xi_a) H ,
$$
after symmetrizing (a,b) the result vanishes. Therefore to
assure gauge invariance we must require (2.20) to vanish, i.e.
$c_{(ab)}=0$. If this condition is satisfied, the dual action
is obtained by adding to $S[\phi,A]$ the Lagrange multiplier
enforcing A to be flat:
\be
{\tilde S}[\phi,A]=S[\phi,A]+\frac{1}{2\pi}\int \chi_a
F_{+-}^a(A).
\ee
 From (2.18,21) we see that the gauge field appears at most
quadratically in the action. We can thus eliminate A by using
its equations of motion up to a jacobian factor which affects
the dilaton transformation rules \cite{r2,r3}.

One of the difficulties in understanding the global properties
of the dual action lies in the fact that apparently there is no
topology in the auxiliary field $\chi$. We will show in the
next few sections that $\chi$ is actually described by a
WZW-model based on the group G used to dualize. To conclude we
remark that if in (2.21) we integrate first over $\chi^a$, we
obtain that $A_{\pm}=g^{-1}\partial_{\pm}g$.
Making a change of variables, as described after equation
(2.4) $\phi^i \rightarrow \phi^{'}\,^i=f^i(\phi,g)$, where $f$
describes the action of the group on the manifold M,
we get back to the original theory.

There is a remarkable identity proved by Hull and Spence
\cite{r11} which will be useful later on. The last three
terms in (2.11) can be expressed in terms of a 3-form.
Define the covariant differentials:
\bea
D\phi^i&\equiv& d\phi^i-\xi_a^i A^a; \nonumber \\
A^a&=& A_{\mu}^a dx^{\mu}; \nonumber \\
F^a&=&dA^a-\frac12 f^a_{bc} A^bA^c;
\eea
Then after some algebra using (2.15,16) the action
becomes:
\bea
4\pi S[\phi,A]&=&\int_{M_2} g_{ij}(\phi)D_{+}\phi^i
D_{-}\phi^j+\int_{B_3}(\frac13 H_{ijk}D\phi^i
D\phi^j D\phi^k+v_{ai}D\phi^iF^a \nonumber \\
&+&c_{(ab)}A^adA^b+
\frac13 f_{mn}^b c_{(ab)}A^mA^nA^a)
\eea
All terms but the last two are gauge invariant, and the
last two are a Chern-Simons term whose gauge variation is
responsible for the anomaly (2.20).

The lesson to be learned from this section is that one way
to think about duality is the equivalence of descriptions
of the same theory based on different gauges. Hence in
trying to carefully construct the quantum properties of the
dual theory we have to understand the interplay between
gauge invariance and conformal invariance. This we do in the
next section.

\section{Mixed anomalies and effective actions}
\setcounter{equation}{0}

To solve the puzzle posed by the example studied in \cite{r7}
we need to analyze carefully the integration measure over the
gauge fields for the action (2.18). The authors of \cite{r7}
considered a four-dimensional Bianchi V cosmological solution
to string theory. The $b_{\mu\nu}$ field vanishes, and the
metric takes the form:
\be
ds^2=-dt^2+a(t)^2(dx^2+e^{-2x}(dy^2+dz^2)).
\ee
Conformal invariance requires $a(t)=t$, in which case the
metric is flat: It is the interior of the light-cone in
Minkowski space. This can be seen by first making the change
of variables $x\rightarrow \ln x$,
$$
ds^2=-dt^2+\frac{t^2}{x^2}(dx^2+dy^2+dz^2),
$$
with $x > 0$, $y,z \in R$. The equal t-surfaces are described
by the upper half-space with constant negative curvature.
By a simple change of variables we can transform the upper
half-space into the mass hyperboloids in Minkowski space.
Hence this theory is indeed conformally invariant.
The group of motions on the equal time surfaces is given by:
\bea
P_0&=&\frac{\partial}{\partial x}+y\frac{\partial}{\partial y}
+z\frac{\partial}{\partial z} \nonumber\\
P_1&=&\frac{\partial}{\partial y}, \,\,\,
P_2=\frac{\partial}{\partial z},
\eea
with commutation relations:
\be
[P_0,P_1]=-P_1,\,\,\, [P_0,P_2]=-P_2,\,\,\,
[P_1,P_2]=0
\ee
The only non-vanishing structure constants are:
\be
f_{01}^1=f_{02}^2=-1
\ee
Since the adjoint representation is given by
$(T_a)_b^c=-f_{ab}^c$, we easily see that $Tr T_0=2$,
$Tr T_1=Tr T_2=0$, therefore the group is not semisimple.
It is worth noting that the group generated by (3.3) must be
a subgroup of the Lorentz group SO(3,1). Indeed, if $K_i$
i=1,2,3 are the boost generators and $J_i$ i=1,2,3 are the
rotation generators, it is easy to see that (3.3) is satisfied
with the identifications:
\be
P_0=K_3,\,\,\, P_1=J_1-K_2,\,\,\, P_2=J_2+K_1.
\ee
If one performs the duality transformation for (3.3) according
to the rules in \cite{r3}, then one finds \cite{r7} that the
dual theory does not satisfy the $\beta$-function equations to
first order in $\alpha^{'}$ no matter what choice one makes for
the dilaton field $\Phi$ \cite{r13}:
\bea
&R_{\alpha\beta}-\frac14 H_{\alpha\beta}^2
-\nabla_{\alpha}\nabla_{\beta}\Phi=0 \nonumber\\
&-\frac12 \nabla^{\gamma}H_{\alpha\beta\gamma}-
\frac12\nabla^{\gamma}\Phi H_{\alpha\beta\gamma}=0
\nonumber\\
&R-\frac{1}{12}H^2-2\nabla^2\Phi-(\nabla\Phi)^2
-2\frac{c-d}{3\alpha^{'}}=0
\eea
where
$$
H_{\alpha\beta\gamma}=\nabla_{[\alpha}
b_{\beta\gamma ]},\,\, H_{\alpha\beta}^2=
H_{\alpha\rho\sigma}H_{\beta}^{\rho\sigma}.
$$
A similar example along the same lines is provided by a
Bianchi IV type cosmological model with metric:
\be
ds^2=-dt^2+a^2(t)dx^2+b^2(t)e^{-x}(dy^2+dz^2).
\ee
Conformal invariance to order $\alpha^{'}$ requires
$a(t)=t/2$, $b(t)=t$; and the group of motions is
generated by
\be
K_1=\frac{\partial}{\partial y},\,\,\, K_2=\frac{\partial}
{\partial z},\,\,\, K_3=\frac{\partial}{\partial x}+
y\frac{\partial}{\partial y},
\ee
with commutation relations
\be
[K_1,K_2]=0,\,\,\, [K_2,K_3]=0, \,\,\,
[K_1,K_3]=K_1.
\ee
Now $Tr K_3^{\rm adj}=1,\,\, Tr K_1^{\rm adj}=
Tr K_2^{\rm adj}=0$, and as in the previous example
the non-abelian dual of (3.7) with respect to (3.9) does not
satisfy the $\beta$-function equations to order $\alpha^{'}$
for any choice of the dilaton field.

In order to understand the origin of the problem we go back
to the general form of the duality transformation. Since we
are interested in conformal invariance, we introduce an
arbitrary metric $h_{\alpha\beta}$ on the world-sheet and
compute the contribution to the trace anomaly of the
auxiliary gauge fields $A_{\pm}^a$. This is also justified
within string theory where in the first quantized formulation
we have to integrate over world-sheet metrics. If for simplicity
we work on genus zero surfaces, the most straightforward way to
compute the dependence of the effective action on the world-sheet
metric is to first parametrize $A_{\pm}$ as:
\be
A_+=L^{-1}\partial_+ L,\,\,\,A_-=R^{-1}\partial_-R,
\ee
for $L,R$ group elements. We can think of $x^{\pm}$ as light-cone
variables or as complex coordinates, and they depend on the
metric being used. In changing variables from $A_{\pm}$ to
$(L,R)$ we encounter jacobians:
\be
{\cal D}A_+{\cal D}A_-={\cal D}L\,{\cal D}R\, det
(D_+(A_+)D_-(A_-))
\ee
with $A_{\pm}$ given by (3.10) (we take $A_{\pm}$ as
antihermitian matrices). We can write the determinants in
(3.11) in terms of a pair of (b,c)-systems ($b_{+a}, c^a$),
($b_{-a}, {\tilde c}^a$). $c,{\tilde c}$ are 0-forms
transforming in the adjoint representation of the group.
For arbitrary groups $b_{\pm a}$ transform in the coadjoint
representation. On a general Riemann surface $\Sigma$, if K
is the canonical line bundle, and E is a vector bundle over
$\Sigma$, $c,{\tilde c}$ are geometrically sections of E,
$c,{\tilde c}\in \Gamma (E)$ for E the adjoint bundle for a
group G, $b_+\in\Gamma (K\otimes E\check{\quad})$,
$b_-\in\Gamma ({\bar K}\otimes E\check{\quad})$, where
$E\check{\quad}$ is the dual
bundle. The determinants in (3.11) can be exponentiated in
terms of the (b,c)- systems with an action:
\be
S[b_{\pm},c,{\tilde c}]=\frac{i}{\pi}\int (b_+D_-(A)c+
b_-D_+(A){\tilde c}),
\ee
which is formally conformal invariant. The variation of S with
respect to the metric is given by the energy-momentum tensor
$T_{\pm\pm}$. We can ignore momentarily that $A_{\pm}$ are
given by (3.10) and work with arbitrary gauge fields.
We can compute the dependence of the effective action for
(3.12) on the metric $h_{ab}$ and the gauge field using
Feynman graphs, operator product expansions (OPE) or heat
kernel methods. Expanding about the flat metric, and using
the methods in \cite{r8}, the first diagrams contributing to
the effective action are

%%Begin InstantTeX Picture
\let\picnaturalsize=Y
\def\picsize{1.0in}
\def\picfilename{figure.eps}
%If you do not have the picture file add:
%\let\nopictures=Y
%to the beginning of the file.
\ifx\nopictures Y\else{\ifx\epsfloaded Y\else\input epsf \fi
\let\epsfloaded=Y
\centerline{\ifx\picnaturalsize N\epsfxsize \picsize\fi
\epsfbox{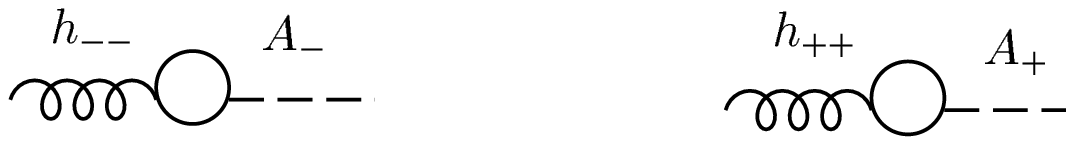}}}\fi
%%End InstantTeX Picture

$h_{--} (h_{++})$ couples to $T_{++} (T_{--})$, and
$A_- (A_+)$ to the ghost currents $j_+ (j_-)$ given by:
\be
T_{++}=\partial_+c^a b_{+a}, \,\,\,
T_{--}=\partial_-{\tilde c}^a b_{-a}
\ee
\be
j_+^i=b_{+a}(T^i)^a_b c^b, \,\,\,
j_-^i=b_{-a}(T^i)^a_b {\tilde c}^b
\ee

If one keeps track of the $i\epsilon$ prescriptions in the
propagators appearing in the graphs, the loop
integrals are finite, and we can write their contributions
to the effective action as:
\be
W^{(2)}=\frac{1}{4\pi}Tr T^a \int d^2p(h_{--}(p)
\frac{p_+^2}{p_-}A_-^a(-p)+h_{++}(p)\frac{p_-^2}{p_+}
A_+^a(-p)).
\ee
The coefficient of (3.15) and $W^{(2)}$ may also be computed
using the OPE:
\be
T(z) j_a(w)=\frac{Tr T_a}{(z-w)^3}+
\frac{1}{(z-w)^2} j_a(w)+
\frac{1}{(z-w)}\partial j_a(w).
\ee
As it stands, $W^{(2)}$ has a gravitational anomaly, i.e. the
energy-momentum tensor is not conserved. However we can still
add local counterterms to (3.15) to recover general coordinate
invariance. Since to first order in $h$ the two-dimensional
scalar curvature has as Fourier transform:
\be
R(p)=2(2p_+p_-h_{+-}(p)-p_+^2h_{--}-p_-^2h_{++}),
\ee
if we add the counterterms:
\bea
W_{\rm c.t.}&=&\frac{1}{4\pi}Tr T_a \int A_-^a(-p)
(h_{++}(p)p_--2p_+h_{+-})\nonumber\\
&+&\frac{1}{4\pi} Tr T_a \int A_+^a(-p)
(h_{--}(p)p_+-2p_-h_{+-}),
\eea
we obtain an effective action
\be
W^{(2)}=\frac{1}{16\pi}Tr T_a \int R(p)
\frac{p_+A_-^a(-p)+p_-A_+^a(-p)}{p_+p_-},
\ee
leading to a conserved energy-momentum tensor,
although it contains a trace anomaly which is not
proportional to $R(p)$ and therefore it cannot be absorbed in
a modification of the dilaton transformation. Varying (3.19)
with respect to $h_{+-}$ leads to:
\be
<T_{+-}>=\frac{\delta W^{(2)}}{\delta h_{+-}}=
\frac{1}{4\pi} Tr T_a (p_+A_-^a(-p)+
p_-A_+^a(-p))
\ee
which in covariant form it becomes $\sim Tr T_a\, \nabla^{\alpha}
A_{\alpha}^a$.

Similarly we can vary the effective action to this order
with respect to gauge transformations to evaluate the corresponding
gauge anomaly:
\bea
&&(D_- \frac{\delta W^{(2)}}{\delta A_-^a}+D_+ \frac{\delta
W^{(2)}}{\delta A_+^a})\sim p_-\frac{\delta W^{(2)}}
{\delta A_-^a(p)}+p_+\frac{\delta W^{(2)}}
{\delta A_+^a(p)} \nonumber\\
&&=p_- <j_{a+}(p)>+p_+<j_{a-}(p)>=-\frac{1}{8\pi}
Tr T_a R(-p).
\eea
This is a different way of writing the third order pole in the
OPE (3.16). This result is not surprising when we recall the
general theory of b-c systems \cite{r14}. The ghost number
current $j_+=bc$ has an anomaly proportional to $R$ (2d scalar
curvature) unless b,c have spin 1/2. In the bosonized form the
b-c system is described by an action of the form
$$
\int (\partial_+\phi\partial_-\phi+(2j-1)R\phi);
$$
so that $\partial^{\mu}j_{(gh)\mu}\sim R$ as in (3.21) and (3.16),
for this reason we cannot gauge the ghost number current unless
other compensating fields are included. In our case we can think
of the generator whose trace is not zero $Tr T_a\neq 0$ as
generating a generalized ghost number current for the auxiliary
b-c systems used to construct the measure over $A_{\pm}$.
Since $b_+$ has spin 1 and c spin 0, we are certain that the
anomaly is there and that there is no consistent gauging,
certainly not compatible with conformal invariance.
 From (3.20) we see that at this order ($W^{(2)}$) the trace
anomaly is not proportional to $R$, and it therefore cannot be
absorbed in a contribution to the dilaton or the effective
value of c (the central charge of the Virasoro algebra).
The contribution in (3.20) spoils the conformal invariance
of the dual theory, and further fields should be required
to cancel it. However in that case the resulting theory
would not agree with the one obtained through a naive
duality transformation.

A third method to obtain the same conclusion as in (3.20) is
to use heat kernel methods. Using now complex variables, the
problem we are trying to address is the computation of the
product of chiral determinants:
\be
det D_{{\bar z}}(A)\, det D_{z}(A)
\ee
Each determinant does not make too much sense by itself,
not only because it is infinite but because both $D_{z}(A)$
and $D_{{\bar z}}(A)$ map very different spaces
$D(A): \Gamma[E]\rightarrow \Gamma[K\otimes E]$, and
${\bar D}(A): \Gamma[E]\rightarrow \Gamma[{\bar K}
\otimes E]$. There is also a natural pairing between
$a\in\Gamma[K\otimes E]$ and $b\in\Gamma[{\bar K}
\otimes E\check{\quad}]$:
\be
\int <a_z,b_{\bar z}>dz\wedge d{\bar z},
\ee
where $<,>$ is the pairing between $E$ and $E\check{\quad}$.
To compute
determinants we need operators which do not change the
space where they act. By looking at the graphical evaluation
of the determinants we learn that one is really computing:
\be
\frac{det\nabla^z D_z(A)\, det\nabla^{\bar z}D_{\bar z}(A)}
{det \nabla^z \nabla_z}.
\ee
This is manifestly general coordinate invariant, and it can be
evaluated with heat-kernel methods. We are in fact
interested in the conformal variation of (3.24). Define
\be
H_+\equiv \nabla^z D_z (A), \,\,\,\,
H_-\equiv \nabla^{\bar z} D_{\bar z}(A),
\ee
Then under a general variation:
\bea
\delta\log det H_{\pm}&=&-\int_{\epsilon}^{\infty}
\frac{ds}{s} Tr s\delta H_{\pm}e^{sH_{\pm}}=
-\int_{\epsilon}^{\infty}ds Tr\, \delta H_{\pm}
e^{sH_{\pm}} \nonumber\\
&=&-\int_{\epsilon}^{\infty}ds \frac{\partial}
{\partial s}Tr \delta H_{\pm}H_{\pm}^{-1}
e^{sH_{\pm}}=Tr \delta H_{\pm}H_{\pm}^{-1}
e^{\epsilon H_{\pm}}
\eea
Under a conformal transformation $\delta
h_{\alpha\beta}=\delta\rho h_{\alpha\beta}$, and
\be
\delta H_{\pm}=-\delta\rho H_{\pm},
\ee
hence:
\be
\delta_{\rho}\log det H_{\pm}=-Tr \delta\rho
e^{\epsilon H_{\pm}}=-\int dx\delta\rho (x)K_{\pm}
(x,x;\epsilon),
\ee
where $K_{\pm}(x,y;\epsilon)$ is the heat kernel of $H_{\pm}$.
To compute the $O(\epsilon^0)$ in (3.28) we can use
Gilkey's results \cite{r15}, or simply take a Fourier transform.
To the order at which we are working there is no difference.
Hence,
$$
\nabla^z D_z(A)=\nabla^z \nabla_z+(\nabla^z A_z)+
A_z\nabla^z ,
$$
and (3.28) becomes:
$$
\delta_{\rho}\log det H_+=-\int dx \delta\rho (x)\int
\frac{d^2 k}{(2\pi)^2}e^{[(ik^z+\nabla^z)(ik_z+\nabla_z)
+\nabla^z A_z+A_z(\nabla^z+ik^z)]}.1 ,
$$
rescaling $k\rightarrow \epsilon^{1/2}k$, noting that
$\int d^2 k e^{-k^2}k^z k^z=\int d^2 k e^{-k^2}
k^{\bar z}k^{\bar z}=0$, and recalling that $k^z k_z=
(k_x^2+k_y^2)/2$, because $h_{z{\bar z}}=1/2$ for the
flat metric, we obtain:
\bea
\delta_{\rho}\log det H_+&=&-\frac{1}{4\pi}\int\delta
\rho Tr \nabla^z A_z \nonumber\\
\delta_{\rho}\log det H_-&=&-\frac{1}{4\pi}\int\delta
\rho Tr \nabla^{\bar z} A_{\bar z},
\eea
as in (3.20), but now the computation is done for the
full determinant and not just to first order in $A$.
Hence the diagrammatic and the heat kernel evaluations
agree, and we conclude that the condition for the duality
transformation to respect conformal invariance is that the
generators of the duality group in the adjoint representation
should have a vanishing trace. Once again this may happen
only for non-semisimple groups, as in the examples discussed
at the beginning of this section. Next we will study the
non-abelian dual of WZW-models for semisimple groups.

\section{Duality in WZW models}
\setcounter{equation}{0}

A very large class of CFT with non-abelian symmetries is
embodied in the WZW models \cite{r9} for any group G which
we will take to be semisimple, and the GKO-constructions
\cite{r6,r16}.

In this section we present the exact non-abelian duality
transformation for a generic WZW-model with group G with
respect to the vector action of any of its subgroups.
Although we will focus our attention on the non-abelian
case, our construction applies as well to abelian, or
mixed transformations. Our procedure allows us to
determinate the global structure of the non-abelian dual,
something which was not yet known. The main result of this
section is to show that if we start with a WZW-model with
group G and level k, its global symmetry group is $G_L\times
G_R$, and $G_D$ (the diagonal embedding of G) is anomaly
free. Hence we can select a subgroup $H\subset G_D$; and
the exact non-abelian dual with respect to H is the coset
model $(G/H)_k$ times a WZW-model for H at level k (if the
Dynkin index of the embedding is one which we take for
simplicity; the extension to other cases is straightforward).
The basic ingredients in the proof are the use of the Polyakov-
Wiegmann (PW) property of the WZW-action together with a
special representation of the auxiliary field $\chi^a$ used
in (2.21) to define the starting point of the duality
transformation. It is the last step which unravels the
global topology hidden in the non-abelian dual.

The WZW-action \cite{r9} for a group G is given by $k S_0[g]$,
\be
-4\pi S_0[g]=\int_{M_2} dx^-dx^+ Tr (g^{-1}\dmas g g^{-1}
\dmenos g)+\frac13 \int_{B_3,\partial B_3=M_2}
Tr (g^{-1}dg)^3,
\ee
where $B_3$ is a three-manifold whose boundary is the space-time
$M_2$ and for compact semisimple groups k is a positive integer.

A basic property of (4.1) is \cite{r17}:
\be
S_0[g_1g_2]=S_0[g_1]+S_0[g_2]-\frac{1}{2\pi}\int
Tr (g_1^{-1}\dmas g_1 \dmenos g_2 g_2^{-1}).
\ee
we next select a subgroup $H\subset G$, and according to the
rules of duality we gauge it. The anomaly free condition
requires that $H$ should be anomaly free (that (2.20) vanishes),
hence we choose the vector action of $h$ on $g$, and at the same
time we introduce the gauge fields $A_{\pm}$ (as antihermitian
matrices in the Lie algebra of $H$). The gauge transformations
are:
\be
g\rightarrow h^{-1} g h, \,\,\, A_{\pm}\rightarrow h^{-1}
(A_{\pm}+\partial_{\pm})h.
\ee
The most straightforward way to gauge (4.1) is to use the
Noether procedure. After some algebra one obtains \cite{r9}
\bea
S[g,A_{\pm}]&=&S_0[g]-\frac{1}{2\pi}\int Tr(A_+\dmenos g g^{-1}
-A_-g^{-1} \dmas g) \nonumber\\
&+&\frac{1}{2\pi}\int Tr(A_+gA_-g^{-1}-A_+A_-).
\eea
If we work on spherical world-sheets, we can rewrite $A_{\pm}$ in
terms of two group elements:
\be
A_+=L^{-1}\dmas L,\,\,\, A_-=R^{-1}\dmenos R.
\ee
Replacing (4.5) in (4.4) and using (4.2) leads to:
\be
S[g,A_{\pm}]=S_0[LgR^{-1}]-S_0[LR^{-1}]
\ee
Since under gauge transformations:
\be
g\rightarrow h^{-1}gh,\,\,\, L\rightarrow Lh,\,\,\,
R\rightarrow Rh,
\ee
we learn that (4.6) is manifestly gauge invariant. In the
standard treatment of duality we add the extra term
$\chi_a F_{+-}^a$ and then eliminate the gauge field.
However, we find it more
convenient to carry the gauge fields along till the end.
This will clarify the global properties of the non-abelian
dual.

The next step is to transform the auxiliary field term into
a more appealing form. We can write $\chi$ and F as
antihermitian matrices valued in the Lie algebra of $H$.
With the representation (4.5):
\bea
F_{+-}&=&\dmas A_--\dmenos A_++[A_+,A_-] \nonumber\\
&=&\partial_+(R^{-1}\partial_-R)-\dmenos (L^{-1}
\dmas L)+[L^{-1}\dmas L,R^{-1}\dmenos R].
\eea
Defining:
\bea
D_+f&\equiv& \dmas f+[L^{-1}\dmas L,f] \nonumber\\
D_-f&\equiv& \dmenos f+[R^{-1}\dmenos R,f],
\eea
and using the identities:
\bea
\dmenos (L^{-1}\dmas L)&=&D_+(L^{-1}\dmenos L) \nonumber\\
\dmas (R^{-1}\dmenos R)&=&D_-(R^{-1}\dmas R),
\eea
we can write the gauge field strengh $F_{+-}$ as
\bea
F_{+-}&=&D_+ (R^{-1}\dmenos R-L^{-1}\dmenos L)\nonumber\\
&=&D_-(R^{-1}\dmas R-L^{-1}\dmas L).
\eea
Integrating by parts (and up to total
derivatives) the auxiliary term becomes:
\bea
-Tr \chi F_{+-}&=&Tr D_+\chi (R^{-1}\dmenos R-L^{-1}\dmenos L)
\nonumber\\
&=&Tr D_-\chi (R^{-1}\dmas R-L^{-1}\dmas L),
\eea
Using the identities:
\bea
D_+\chi &=&L^{-1}\dmas (L\chi L^{-1}) L \nonumber\\
D_-\chi &=&R^{-1}\dmenos (R\chi R^{-1})R
\eea
and defining the variable
\be
\Sigma=LR^{-1},
\ee
we obtain:
\bea
Tr \chi F_{+-}&=&-Tr \dmas (L\chi L^{-1})\Sigma \dmenos
\Sigma^{-1} + {\rm total \, deriv.} \nonumber\\
&=&Tr \dmenos (R\chi R^{-1})\Sigma^{-1}\dmas\Sigma +
{\rm total \, deriv.}
\eea
Collecting our results, the starting point for duality is
\bea
&&S[g,A_{\pm}]+\frac{1}{2\pi}\int Tr \chi F_{+-}=
kS_0[LgR^{-1}]-kS_0[\Sigma] \nonumber\\
&&+\frac{1}{2\pi}\int Tr
\dmenos (R\chi R^{-1})\Sigma^{-1}\dmas\Sigma.
\eea
And the quantum theory is defined by integrating over
$(A_{\pm},g,\chi)$. Since we find it preferably to work with
$L,R$-fields, we need to evaluate the jacobian in the change of
variables from $A_{\pm}$ to ($L,R$).Under infinitesimal
changes of $L,R$ $A_{\pm}$ change according to:
\bea
\delta A_+&=&\dmas (L^{-1}\delta L)+[A_+,L^{-1}\delta L] \nonumber\\
\delta A_-&=&\dmenos (R^{-1}\delta R)+[A_-,R^{-1}\delta R],
\eea
and the change of variables is:
\be
{\cal D} A_+ {\cal D} A_-=det D_+(A_+) det D_-(A_-)\,
{\cal D}L {\cal D}R,
\ee
where the determinants are in the adjoint representation for
$A_{\pm}$ as in (4.5). These determinants can be computed by
integrating the two-dimensional gauge anomaly \cite{r17}.
The result is:
\be
{\cal D}A_+ {\cal D}A_-={\cal D}L {\cal D}R
e^{-i c_H S_0[\Sigma]} (det \dmas
det \dmenos)^{{\rm dim} H},
\ee
where a counterterm $\int A_+ A_-$ has been added to preserve
vector gauge invariance. $c_H$ is the second Casimir of $H$ in the
adjoint representation, and it appears because we write the
WZW-action $S_0[\Sigma]$ in the fundamental representation.
Exponentiating the determinants in (4.19) using two pairs of
ghosts $(b_{a+},c^a)$, $(b_{a-},{\tilde c}^a)$, we obtain the
total action:
\bea
&&k S_0[LgR^{-1}]-(k+c_H)S_0[\Sigma]+\frac{1}{2\pi}\int
Tr \dmenos (R\chi R^{-1})\Sigma^{-1}\dmas \Sigma \nonumber\\
&&-\int (b_{a+}\dmenos c^a+b_{a-}\dmas {\tilde c}^a)
\eea
integrated over $(g,L,R,\chi,b_{\pm},c,{\tilde c})$.
The first, second and fourth terms in (4.20) look
essentially like the functional integral representation
for the GKO-construction of the $(G/H)$ coset conformal
field theory \cite{r18}. To finally conclude that this is
indeed part of the answer we still have to treat the
auxiliary field term. However a comparison of (4.20) and
(4.1) immediately suggests the following procedure. First
introduce the gauge invariant variables
\be
\chi_R=R\chi R^{-1},\,\,\, \chi_L=L\chi L^{-1},
\ee
next we can think of $\dmenos\chi_R$ as a gauge connection.
On the sphere we can write this gauge connection as a pure
gauge field (as in (4.5)):
\be
\dmenos\chi_R=(k+c_H)\beta^{-1}\dmenos\beta,
\ee
where $\beta$ takes values on $H$ and the normalization factor
($k+c_H$) is chosen for convenience. In computing the jacobian
factor between $\chi_R$ and $\beta$ we proceed in two steps,
first we change from $\chi_R$ to $\dmenos\chi_R/(k+c_H)$, this gives
a jacobian equal to $(det \dmenos)^{-{\rm dim}H}$; then we
compute the change between $\dmenos\chi_R/(k+c_H)$ and
$\beta^{-1}\dmenos
\beta$. This however is the same computation as the one leading
to (4.19) and it is carried out by integrating the anomaly.
Thus:
\be
{\cal D}\chi=({\cal D}\beta)\exp{(-i c_H)}
S_0[\beta^{-1}].
\ee
The $(det \dmenos)$ factors cancel out and there is no need to
introduce extra ghosts. Since everything is expressed in terms
of gauge invariant variables, we can factor out the volume of
the gauge group and integrate only over the combination
$\Sigma=LR^{-1}$. With all the previous manipulations we have
brought the partition function to the form:
\bea
&&\int {\cal D}g {\cal D}\Sigma {\cal D}\beta {\cal D}b
{\cal D}c \exp i(k S_0[LgR^{-1}]-(k+c_H)S_0[\Sigma] \nonumber\\
&&-\frac{k+c_H}{2\pi}\int Tr \beta^{-1}\dmenos
\beta\Sigma^{-1}\dmas\Sigma -c_H S_0[\beta^{-1}]+
{\rm ghosts}).
\eea
Using (4.2) once again leads to:
\bea
&&\int {\cal D}g {\cal D}\Sigma {\cal D}\beta {\cal D}b
{\cal D}c \exp i(k S_0[LgR^{-1}]-(k+c_H)S_0[\Sigma\beta^{-1}]+
kS_0[\beta^{-1}]\nonumber\\
&&-\int (b_{a+}\dmenos c^a+b_{a-}\dmas {\tilde c}^a)).
\eea
Now we can make the change of variables $g\rightarrow LgR^{-1}$,
$\Sigma\rightarrow\Sigma\beta^{-1}$ with unit jacobian.
This leads to an action:
\be
kS_0[g]-(k+c_H)S_0[\Sigma]-\int (b_{a+}\dmenos c^a+
b_{a-}\dmas {\tilde c}^a)+k S_0[\beta^{-1}].
\ee
Although the action looks decoupled, $(g,\Sigma,b,c)$ are coupled
through the physical state condition. Since we have gauged $H$,
there is BRST charge receiving contributions from $g,\Sigma,
b,c$ given by \cite{r18}:
\be
Q=\oint c^a (J_{+a}(g)+J_{+a}(\Sigma)-\frac12c^mf_{am}^n
b_{+n}),
\ee
where $J_{+a}(g)$ are the left-handed $H$-currents from $kS_0[g]$,
$J_{+a}(\Sigma)$ the left-$H$ currents from $-(k+c_H) S_0[\Sigma]$
and the third contribution comes from the ghosts.
It is easy to check that
\be
\{Q,b_{+a}\}=J_{+a}(g)+J_{+a}(\Sigma)+J_{+a}^{{\rm ghost}},
\ee
then $Q^2=0$ as long as the total current has vanishing central
charge. This is indeed the case because $J_+(g)$ has central
charge k, $J_+(\Sigma)$ has $-(k+c_H)$ and $J_+^{gh}$ $+c_H$.
Hence the first three terms in (4.26) together with the physical
state condition $Q|phys>=0$ provide a functional integral
representation of the coset conformal field theory $G/H$.
Since the energy-momentum tensor associated to WZW-models
is of Sugawara form, the total central charge of the Virasoro
algebra coming from the first three terms is easily
computed once we use the relation noted in Karabali et al
\cite{r18}:
\be
c(H,-k-c_H)-2 dim H=-c(H,k).
\ee
The contribution $-2 {\rm dim}\,H$ comes from the b-c system, and as
mentioned before we are assuming that the embedding $H\subset G_D$
has Dynkin index equal to 1, otherwise there are obvious changes
in the normalization factors for the WZW-models for the subgroup
H with respect to that of G (and this also affects
$S_0[\beta^{-1}]$).
Hence although we have not explicitly eliminated the gauge field
variables $\Sigma$ in the classical Lagrangian (4.26),
it is clear that (4.26) describes the non-abelian dual.
What is more important is that (4.26) brings out the global
topology of the auxiliary field $\chi$ once it is represented
in terms of $\beta$. Although the relation between $\chi_R$
and $\beta$ is non-local, the Lagrangians (4.26), (4.24) are
local and we can study their quantum theory with standard
methods.

If we want to explicitly eliminate the field $\Sigma$ altogether,
we may proceed as follows. Since $H\subset G_D$, we may
orthogonally decompose the Lie algebra of G $\underline{g}=
\underline{h}\oplus \underline{k}$,
where $[\underline{h},\underline{k}]\subset \underline{k}$.
Then the element $g\in G$ can be
written as $g=lh$, with $h\in H$, and $l^{-1}dl\in \underline{k}$.
Since the decomposition $\underline{h}\oplus \underline{k}$
is orthogonal,
\be
S_0[lh]=S_0[l]+S_0[h],
\ee
with the last term in (4.2) equal to zero due to orthogonality.
Hence the action takes the form
\be
S_{{\rm dual}}=k S_0[l]+k S_0[\beta^{-1}]+
(k S_0[h]-(k+c_H)S_0[\Sigma]+{\rm ghost}).
\ee

The terms between parenthesis represent a CFT with $c=0$,
and the BRST conditions only involve the $H$-valued fields
$h,\Sigma$ and the ghosts $b_{\pm},c,{\tilde c}$.
Hence these terms represent an ($H/H$) theory with $c=0$
whose Hilbert space is reduced to the identity operator.
Therefore we can set the integral over
$(h,\Sigma,b_{\pm},c,{\tilde c})$ (taking into account the
BRST condition) to 1, and write the dual action explicitly
in terms of the coset variables $l$ together with a level k
WZW-model for the $H$-valued auxiliary field. This completes the
proof of the main statement at the beginning of the section.
We now present some consequences and remarks concerning the
results presented.

It should be clear from the previous arguments that if we
consider a coset CFT $G/H$ such that the centralizer $H^{'}$
of $H$ in $G$ is not trivial, we can perform duality with
respect to $H^{'}$, and the result will be (4.26) with two
extra terms, $-(k+c_{H^{'}}) S[\Sigma^{'}]+
S_{\rm gh}[b^{'},c^{'}]$ associated to $H^{'}$.

Another simple consequence of our arguments is that they can
readily be specialized to the case when $H$ is abelian.
In this case the dual CFT is given in terms of $G/H$ parafermions
and a collection of free fields living on the torus defined
by $H$, as one expect from standard arguments of abelian duality
(see for example the review by E. Kiritsis in [2]).

We remark further that we have not written an explicit form
for the metric induced in $G/H$ by our construction.
This is reasonable because we have carried out an exact
quantum treatment. In the limit as $k\rightarrow\infty$ the
leading form of the metric is given by the one obtained from
the left hand side of (4.16) after eliminating the gauge fields
$A_{\pm}$ and then fixing the gauge. However we expect
corrections to all orders in 1/k, as exemplified by the
corrections to the two-dimensional String Black Hole
\cite{r19} described in \cite{r20}. There are in principle
methods to compute systematically the 1/k corrections to
the classical metric \cite{r21}. The advantage of working
without explicitly using the metric is that the classical
metric for the dual theory with respect to $H\subset G$ is
quite singular, there are fixed submanifolds of $G$ under the
action of $H$, thus leading to nasty singularities in the
naive metric; and this makes it nearly hopeless to obtain
(4.31) in the standard $\sigma$-model approach.

One important point in our derivation of (4.31) was the relative
normalization between $\chi_R$ and $\beta$ (the factor $k+c_H$).
In the original approach, the normalization of $\chi$ was
not fixed. In the abelian case the range of the auxiliary
field $\chi$ (which for compact abelian $H$ is a torus) is
determined by requiring the absence of modular anomalies for
world-sheets with genus higher than zero (see for instance
\cite{r5}, \cite{r4} and \cite{r22}).
This related the size of the orbits of the isometry group
with the size of the dual torus where the auxiliary field
takes its values. More explicitly this is a consequence of
requiring that the topological term in the second line in
(2.10) should have integral values upon integration
over the world-sheet. The analysis of modular invariance in
genus one is more complicated for (4.26,31), but it is certainly
necessary to determine the detailed operator mapping.
The partition function at genus one of the original WZW-model is
not a product of any modular invariant partition function for
the $G/H$ coset theory and one of the $H$ WZW-model. The Kac-Moody
characters of the $G_k$ WZW-model have a well-defined
decomposition in terms of products of $G/H$ and $H_k$
characters. We have seen that the relative normalization in
(4.22) is justified by the matching of the Virasoro central
extension for the $G_k$ model and its dual. Further justification
comes from the analysis on the torus concerning modular invariance.
Now it is not sufficient to write $A_+=L^{-1}\dmas L$,
$A_-=R^{-1}\dmenos R$, but one has to include the contribution
from flat connections. The arguments follow closely the
evaluation of $G/H$ character in the work of Gawedzki and
Kupiainen \cite{r18}. The details will be presented elsewhere.

Summarizing the field $\beta$ related to $\chi_R$ by (4.22) is
indeed valued in the group manifold $H$; and this determines the
global properties of the non-abelian dual.

It is worth noting to conclude this section that the fields
$\chi_L$, $\chi_R$ resemble the non-local fields used in the
standard Kramers-Wannier duality transformation \cite{r1}.
For instance $\chi_R$ is constructed by multiplying $\chi (x)$
by the Wilson line for $A_-$ from $-\infty$ to x on the left,
and by the $A_-$ Wilson line from x to $\infty$ on the right.
 From this point of view it is also satisfying that the physical
operators in the dual theory (4.26,31) contain contributions
from $\beta$ which are an analogue of the disorder operator
in the Ising model.

\section{General $\sigma$-models with chiral currents}
\setcounter{equation}{0}

Using the results of Hull and Spence \cite{r11} quoted in
section 2 we want to study the general case of a $\sigma$-
model where there are left and right-chiral currents. We will also
present a ``heterotic'' generalization of the treatment of
WZW-models carried out in the previous section. The aim of
this section is to prove that for a general $\sigma$-model
with $G_L\otimes G_R$ isometry group acting on a manifold M,
the non-abelian dual with respect to $G_D\subset G_L\otimes G_R$
is again a generalized GKO construction (M/G) times a WZW-model
for G. To make the arguments as clear as possible, consider
first a model with only left chiral currents:
$\dmenos J_{+a}=0$. From the equations (2.11-21), we have
$v_{ai}=-\xi_{ai}$, so that the action becomes:
\be
4\pi S[\phi,A_-]=\int_{M_2}g_{ij}\dmas\phi^i\dmenos\phi^j
+\frac13 \int_{B_3,\partial B_3=M_2} H_{ijk}
d\phi^i d\phi^j d\phi^k-2\int A_-^a \xi_{ai}^L
\dmas\phi^i
\ee
This model has an anomalous variation:
\be
4\pi \delta S[\phi,A]=-2\int A_-^a \xi_a.\xi_b
\dmas \epsilon^b.
\ee
$\xi_a.\xi_b$ are constants independent of $\phi$ (see the proof
after equation (2.20) when $v_{ai}=\pm\xi_{ai}$).
Note that only $A_-^a$ appears in (5.1). Hence we can take
$A_-=L^{-1}\dmenos L$, and in fact define also $A_+=L^{-1}
\dmas L$, as though A were pure gauge. By adding and
substracting appropriate terms in (5.1), we obtain:
\bea
4\pi S[\phi,A]&=&\int g_{ij}D_+ \phi^i D_- \phi^j+
\frac13 \int_{B_3} H_{ijk}d\phi^i d\phi^j d\phi^k \nonumber\\
&+&\int (A_+^a \xi_{ai}^L\dmenos \phi^i-A_-^a \xi_{ai}^L
\dmas\phi^i)-\int \xi_a.\xi_b A_+^a A_-^b
\eea
where $D_{\pm}\phi^i=\partial_{\pm}\phi^i-\xi_{a}^i
A_{\pm}^a$. Next, we can use the identity (2.23) \cite{r11}
to rewrite (5.3) as:
\bea
4\pi S[\phi,A]&=&\int g_{ij}D_+ \phi^i D_- \phi^j+
\frac13 \int_{B_3} H_{ijk}D\phi^i D\phi^j D\phi^k \nonumber\\
&-&k\int_{B_3} (A^a dA^a+\frac13 f_{abc} A^aA^bA^c)-
k\int \delta_{ab}A_+^a A_-^b,
\eea
since we can take the gauge one-form A as pure gauge
$A=L^{-1}dL$, the curvature term $F^a$ in (2.23) drops
out. Also using the fact that $\xi_a.\xi_b$ is a constant
symmetric metric, we have chosen a basis to diagonalize it.
The normalization factor k depends on the normalization
of the three form H, and it is an integer for compact
semisimple groups. For $A=L^{-1}dL$, the Chern-Simons
term becomes:
\be
k \int (A^a dA^a+\frac13 f_{abc}A^aA^bA^c)=
-\frac{k}{3}\int_{B_3} Tr (L^{-1}dL)^3,
\ee
and consequently the last two terms in (5.4) produce a
WZW-action for G at level k. The first two terms can be
written now in terms of the original theory. The group G
acts on the manifold through some functions
$f: G\times M\rightarrow M$, $\phi\rightarrow f(\phi;L)$.
Using the invariance of $g_{ij}$ and $H_{ijk}$ under
the action of the group, we can make the change of variables
$\phi^i\rightarrow \phi^{'}\,^i=f^i (\phi;L(x))$, with this
change of variables the first two terms in (5.4) become the
original ungauged action written in terms of $\phi^{'}$.
Hence:
\be
S[\phi,A]=\frac{1}{4\pi}\int g_{ij}(\phi^{'})
\dmas\phi^{'}\,^i\dmenos\phi^{'}\,^j+
\frac{1}{12\pi}\int H_{ijk}(\phi^{'})d\phi^{'}\,^i
d\phi^{'}\,^jd\phi^{'}\,^k-kS_0^{{\rm WZW}}[L]
\ee
Thus we decouple the $\sigma$-model variables and the gauge
variables. The model (5.6) is anomalous, and therefore the
standard duality transformation does not work. We can
nevertheless generalize the duality transformation to this
case by analogy with the similar situation in the abelian
case (see first entry in \cite{r4}). We cancel the anomaly in (5.6)
by giving special gauge transformation properties to the
auxiliary field $\beta$ (see (4.16)-(4.22)).
If we want to perform duality with respect to the left-acting
group $G_L$, we only need to integrate over $A_-$.
Since the field $A_+$ is not present, we write the auxiliary
field term in the form (4.20)
\be
\frac{1}{2\pi}\int Tr\dmenos\chi L^{-1}\dmas L
\ee
Making the change of variables $A_-\rightarrow L$ and as
in (4.22) $(k+c_G)\dmas\chi=-\beta^{-1}\dmas\beta$,
together with the Polyakov-Wiegmann property (4.2), we
arrive at
\bea
&&\frac{1}{4\pi}\int g_{ij}(\phi^{'})\dmas\phi^{'}\,^i
\dmenos\phi^{'}\,^j+\frac{1}{12\pi}\int H_{ijk}
(\phi^{'})d\phi^{'}\,^i d\phi^{'}\,^j d\phi^{'}\,^k\nonumber\\
&&-(k+c_G)S_0^{{\rm WZW}}[L\beta^{-1}]+k S_0[\beta^{-1}]+
{\rm ghosts}(b,c)
\eea
This is again interpreted as a coset construction ($M/G_L$)
times a WZW-model for $G_L$ at level k. Note that the term
(5.7) (if written in terms of $\beta$) is precisely what
is needed to cancel the gauge anomaly in (5.6).
As a further remark, note that the original gauged action
(5.3) together with (5.7) are equivalent to the ungauged
theory (5.1) with $A_-=0$. This is because the equations of
motion derived from (5.7) imply that we can choose the gauge
transformation L as a function of $x^+$ only.
But since the original theory has left-chiral invariance,
a simple change of variables recovers the starting action.
This construction can be generalized to perform non-abelian
duality with respect to separate groups on the left and on
the right, provided we include two auxiliary fields
$\beta_L,\beta_R$ needed to cancel the anomaly.
This should have interesting consequences for heterotic
$\sigma$-models, which are currently under investigation.

The case with left- and right-handed currents parallels
the previous arguments and those of section 4.
Once again it is crucial to use the identity (2.23).
Imagine we start with a general WZW-model
\be
4\pi S_0[\phi]=\int g_{ij}(\phi)\dmas\phi^i
\dmenos\phi^j+\frac13\int H_{ijk}d\phi^id\phi^j
d\phi^k,
\ee
with left-handed and right-handed isometries. Then, under
\be
\delta\phi^i=\epsilon^a_L(x)\xi_{La}^i-\epsilon^a_{R}(x)
\xi_{Ra}^i,
\ee
(5.9) changes to
\be
4\pi \delta S_0[\phi]=2\int (\xi_{aLi}\dmas\phi^i
\dmenos\epsilon_{L}^a-\xi_{aRi}\dmenos\phi^i
\dmas\epsilon_R^a)
\ee
The left- and right-Killing vectors satisfy the commutation
relations:
\bea
&&[\xi_{La},\xi_{Lb}]=f_{ab}^{c}\xi_{Lc} \nonumber\\
&&[\xi_{Ra},\xi_{Rb}]=-f_{ab}^{c}\xi_{Rc} \nonumber\\
&&[\xi_{La},\xi_{Rb}]=0
\eea
To cancel (5.11) we introduce:
\be
4\pi S_1[\phi,A]=-2\int (\xi_{aLi}\dmas\phi^i A_{L-}^a-
\xi_{aRi}\dmenos\phi^i A_{R+}^a)
\ee
One can continue like this until we obtain an action whose
variation is independent of $\phi$. If we continue gauging
the left and right isometries, we of course obtain an
anomalous theory. Then we have two options, one is to
follow the heterotic path explained before, and the second is
to restrict the gauging to the diagonal, anomaly free subgroup.
We will follow the latter. Since
\be
\delta A_{L,R}^a=\partial\epsilon_{L,R}^a+f^a_{bc}
A^b_{L,R}\,\epsilon^c_{L,R},
\ee
take now:
\be
\epsilon_L^a=\epsilon_R^a=\epsilon^a.
\ee
Choosing a basis such that:
\be
\xi_{La}.\xi_{Lb}=\xi_{Ra}.\xi_{Rb}=k\delta_{ab},
\ee
we find that the following action is vector-gauge invariant:
\bea
4\pi S[\phi,A_L,A_R]&=&\int g_{ij}(\phi)\dmas\phi^i
\dmenos\phi^j+\frac13\int H_{ijk}d\phi^id\phi^j
d\phi^k \nonumber\\
&-&2\int (\xi_{aLi}\dmas\phi^iA_{L-}^a-\xi_{aRi}
\dmenos\phi^iA_{R+}^a) \nonumber\\
&+&2\int (\xi_{aL}.\xi_{bL}-\xi_{aL}.\xi_{bR})
A_{L-}^aA_{R+}^b.
\eea
One important property of (5.17) is that $A_L^a$ ($A_R^a$)
only appears in the Lagrangian through its $A_{L-}^a
(A_{R+}^a)$ component. Thus we can take $A_L=L^{-1}dL$,
$A_R=R^{-1}dR$, although only the - and + respectively
components do appear. This is useful when we use the identity
(2.11) to write the final form of the action. Since the
full covariant derivatives are:
\be
D_{\pm}\phi^i=\partial_{\pm}\phi^i-\xi_{La}^i A_{L\pm}^a
+\xi_{Ra}^i A_{R\pm}^a,
\ee
as we did in the chiral case we can add and substract some
terms to (5.17) to complete the covariant derivatives.
The result is:
\bea
&&4\pi S[\phi,A_L,A_R]=\int g_{ij}D_+\phi^iD_-\phi^j+
\frac13\int H_{ijk}d\phi^id\phi^jd\phi^k \nonumber\\
&&+\int (\xi_{aLi}d\phi^i\wedge A_L^a+
\xi_{aRi}d\phi^i\wedge A_R^a)+
\int \xi_{aL}.\xi_{bR}(A_{L+}^a A_{R-}^b-A_{L-}^a
A_{R+}^b) \nonumber\\
&&-\int (\xi_{La}.\xi_{Lb}\, A_{L+}^a A_{L-}^b+
\xi_{Ra}.\xi_{Rb}\, A_{R+}^a A_{R-}^b)+
2\int \xi_{aL}.\xi_{bL}\,A_{L-}^a A_{R+}^b.
\eea
The first four terms on the right-hand side
of (5.19) can be combined using (2.23) to produce
$\frac13 H_{ijk}D\phi^iD\phi^jD\phi^k$ on the three-dimensional
integral ($D\phi^i=d\phi^i-\xi_{La}^i A_L^a+\xi_{Ra}^i A_R^a$),
together with the Chern-Simons terms for $A_L$ and $A_R$.
Since $A_L=L^{-1}dL$, $A_R=R^{-1}dR$, these together with the last
two terms in (5.19) reconstruct the WZW-action for G at level-k,
where k is the factor in (5.16). Summarizing we obtain:
\be
4\pi S[\phi,A_L,A_R]=\int g_{ij}D_+\phi^iD_-\phi^j+
\frac13\int H_{ijk}D\phi^iD\phi^jD\phi^k-kS_0^{{\rm WZW}}
[LR^{-1}]
\ee
Now since the group $G_L\otimes G_R$ acts on M, we can make
again the change of variables $\phi^i\rightarrow \phi^{'}\,^i=
f^i(\phi;L,R)$, and the invariance of $g_{ij}$, $H_{ijk}$
imply that
\be
4\pi S[\phi,A_L,A_R]=4\pi S[\phi^{'},0,0]-
kS_0^{{\rm WZW}}[LR^{-1}],
\ee
where $S_0^{{\rm WZW}}[LR^{-1}]$ is the WZW-action (4.1).
Once $\phi^{'},L,R$ are decoupled, the rest of the argument is
a rerun of the WZW case considered in the previous section and
we omit it. This proves the statement at the beginning of this
section.

In the case when the isometries are not chiral (i.e. when we
have the gauge Lagrangian (2.18,23)), the treatment for the
measure for $A_+$, $A_-$ and the change of variables from
$\chi$ to $\beta$ can be done along the same lines; however
one does not have a neat decoupling between the $\phi^i$
variables and $(L,R,\beta)$, for this reason it is not clear
what is the correct normalization factor $\lambda \dmenos\chi=
\beta^{-1}\dmenos\beta$. As in the WZW case $\lambda$ should be
determined in order to match the central charge of the initial
and final conformal field theories, but we do not yet have a
general procedure to compute $\lambda$ explicitly. Note that
$\lambda$ is crucial in order to determine the global topology
of the auxiliary variable $\beta$ in the dual theory. Thus the
exact form of the transformation in the general case when there
are no chiral currents remains open. One extreme case corresponds
to $\sigma$-models with $H=0$. There we have to study each model
separately.

In the next section we analyze in detail two simple
examples based on the Eguchi-Hanson and Taub-Nut gravitational
instantons \cite{r10} which have $SU(2)_L\times U(1)_R$ as
isometry groups.

\section{Duals of gravitational instantons}
\setcounter{equation}{0}

Some interesting examples of non-abelian duality transformations
have been already worked out in previous papers \cite{r4},\cite{givroc},
including WZW-models with $G=SU(2)$ (or $SL(2,R)$) with respect
to its own left action (getting in the latter case a
three-dimensional
black hole).

We would like to examine here another interesting class of
examples, namely the asymptotically locally Euclidean self-dual
solutions to Euclidean gravity (gravitational instantons).
All known $SU(2)$-symmetric solutions \cite{r26} fall within the
framework of a
Bianchi IX metric:
\be
ds^2=f^2(r)dr^2+\sum_{i=1}^3 a_l^2(r) (Tr \frac{i\sigma_l}
{2}(g^{-1}dg))^2
\ee
The elements of $SU(2)$ are here parametrized by their Euler
angles,
\be
g=e^{i\frac{\phi}{2}\sigma_3}e^{i\frac{\theta}{2}\sigma_2}
e^{i\frac{\psi}{2}\sigma_3}
\ee
where $0\leq\theta\leq\pi$, $0\leq\phi\leq 2\pi$,
$0\leq\psi\leq 4\pi$.
In this form all these metrics
are invariant under the left action of $SU(2): g\rightarrow
hg$ (that is, the induced action on the Euler angles
parametrizing a sphere $S^3$).

The celebrated Eguchi-Hanson (EH) metric is a particular case
corresponding
to
\be
f^2=\frac{1}{1-(\frac{a}{r})^4},\,\,\, a_1^2=a_2^2=r^2,
\,\,\, a_3^2=r^2(1-(\frac{a}{r})^4)
\ee
and the Euclidean Taub-Nut (ETN) solution to
\be
f^2=\frac14 \frac{r+m}{r-m},\,\,\, a_1^2=a_2^2=r^2-m^2,
\,\,\, a_3^2=m^2\frac{r-m}{r+m}
\ee
We can treat both cases  (as well as the, less interesting from
our point
of view, Fubini-Study metric on $P_2({\cal C})$) simultaneously,
by considering
the two-dimensional
$\sigma$-model corresponding to the general Bianchi IX ansatz,
namely:
\be
S=\frac{1}{2\pi}\int (f^2(r)\partial r{\bar\partial}r+
\sum_{l=1}^3 a_l^2 (Tr \frac{i\sigma_l}{2}g^{-1}\partial g)
(Tr \frac{i\sigma_l}{2}g^{-1}{\bar\partial} g))
\ee
We would like to consider the left action of the full
$SU(2)$ group (with
respect to which the two-dimensional field $r$ is inert), and
work in the unitary gauge, $g=1$. This yields after some standard
manipulations:
\be
{\tilde S}=\frac{1}{2\pi}\int f^2(r)\partial r
{\bar \partial}r+\frac{4}{\Delta (r,\chi)}(16\chi_i\chi_j+
4\epsilon_{ijk}\chi_ka_k^2+\delta_{ij}\frac{a_1^2a_2^2a_3^2}
{a_i^2})\partial\chi_i {\bar \partial}\chi_j
\ee
where $\Delta (r,\chi)\equiv a_1^2a_2^2a_3^2+16 \sum a_i^2
\chi_i^2$ and a dual dilaton given by \cite{r3}
\be
{\tilde \phi}=\log\frac{\Delta}{64}
\ee
Please note the similarities of these formulas with (7.8) of
first reference in \cite{r4}. We also find, like we did there,
that there is both a dual metric and a dual torsion.
To be specific, the dual to the Eguchi-Hanson solution is:
\bea
d{\tilde s}_{EH}^2&=&\frac{4r^4}{r^4-a^4}dr^2+\frac{r^2}
{(r^4-a^4)(r^4+\xi^2)+r^4\rho^2}((\rho^2+r^4-a^4)
d\rho^2\nonumber\\
&+&(r^4+\xi^2)d\xi^2
+2\rho\xi d\rho d\xi+
(r^4-a^4)\rho^2d\phi^2)
\eea
where we have introduced cylindrical coordinates in the
space of the Lagrange multipliers
\be
\chi_1=\rho\cos\phi,\,\,\, \chi_2=\rho\sin\phi,\,\,\,
\chi_3=\xi
\ee
and rescaled $r$ and $a$ by a factor of 2.

The dilaton is given by
\be
{\tilde \phi}_{EH}=\log{(\frac{(r^4-a^4)(r^4+\xi^2)+
r^4\rho^2}{r^2})}
\ee
and the torsion is
\be
{\tilde b}_{EH}=\frac{\rho}{(r^4-a^4)(r^4+\xi^2)+r^4\rho^2}
(\xi (r^4-a^4)d\rho\wedge d\phi+r^4\rho d\phi\wedge d\xi)
\ee

The dual metric is asymptotically flat as
$r\rightarrow\infty$ since it can be checked that the
Riemann tensor vanishes in this limit.

It (like the original EH metric) also has an apparent
singularity at $r=a$, which can also be resolved as a bolt
(without any need to change the periodicity of our coordinates).
The dual metric is not self-dual (it is not Ricci-flat); an
amusing point is, however, that the corresponding string metric
$d{\tilde s}_{SEH}^2\equiv e^{{\tilde \phi}}d{\tilde s}_{EH}^2$
is not only self-dual, but absolutely flat; the corresponding
Riemann tensor is zero.

The other interesting example is the ETN metric, whose dual
reads
\bea
&&d{\tilde s}_{ETN}^2 = \frac{r+m}{r-m}dr^2+
\frac{1}{\Sigma (r,\rho,\xi)}
((\rho^2+4m^2(r-m)^2)d\rho^2 \nonumber\\
&&+(\xi^2+(r^2-m^2)^2)d\xi^2
+2\rho\xi d\rho d\xi+4m^2 (r-m)^2 \rho^2 d\phi^2)
\eea
where $\Sigma (r,\rho,\xi)\equiv \frac{r-m}{r+m}[4m^2(\xi^2+
(r^2-m^2)^2)+\rho^2(r+m)^2]$.

The dilaton is given by ${\tilde \phi}_{ETN}=\log\Sigma$
and the
torsion reads
\be
{\tilde b}_{ETN}=\frac{\rho (r-m)}{r+m}
\frac{1}{\Sigma (r,\rho,\xi)}
(4\xi m^2 d\rho\wedge d\phi+\rho (r+m)^2 d\phi\wedge d\xi)
\ee

The scalar curvature tends to a positive constant $(5/2m^2)$
asymptotically as $r\rightarrow\infty$. In this case, however,
the singularity at $r=m$ is not apparent, but a true curvature
singularity (${\tilde R}\sim \frac{3}{4m}\frac{1}{r-m}$).
The different behavior in this case as compared with
$d{\tilde s}_{EH}^2$ stems from the fact that the norms of the
killing vectors
vanish when $r=m$, which means that the left action
of $SU(2)$ has fixed points in the four-dimensional
manifold precisely at $r=m$.

As a matter of principle, it is interesting to know whether there
are self-dual
solutions to the beta function equations of conformal
invariance of
Callan et al, with a non-trivial dilaton field.
In the strictest sense of the word, the only possibility
allowed is a Ricci-flat
spacetime. Solutions of the beta-function equations with these
characteristics
can be obtained with zero torsion, and linear dilaton field.

There is a possible avenue for a further generalization.
It is well known
that one can recover {\it all} the beta-function equations
(with frozen dilaton)
by considering the Ricci flatness condition, but with respect
to the generalized
connection $\omega^a\,_b =\gamma^a\,_b +h^a\,_b$, where
$\gamma$ is the usual Levi-civita
connection, and $h$ is the connection giving rise to the
torsion $H$. (The Ricci tensor is now non-symmetric,
so both the symmetric and the antisymmetric parts must
be zero; this is why we get two equations from a
single condition).
A natural conjecture is that there might exist a rescaling
of the vierbein (and perhaps of the torsion as well)
such that the Ricci flatness condition of the corresponding
rescaled torsion conveys the beta function equations,
with a dynamical dilaton.

An easy calculation, however, shows this not to be the case, because
one can not get rid of all terms containing derivatives of
the dilaton field.

\section{Conclusions and outlook}
In this paper we have presented a general analysis of non-abelian
duality. We have shown that there is a potential obstruction to
conformal invariance for non-semisimple groups. This obstruction
is embodied in a mixed gauge and gravitational anomaly appearing
in the measure of integration over the gauge fields. This solves
the puzzle raised in \cite{r7} which was one of the motivations
for our work. We then presented the exact form of the non-abelian
duality transformation for WZW-models with respect to any of its
anomaly free subgroups, and we extended the construction to coset
conformal field theories and to conformal theories on manifolds
with left and right chiral currents. Our treatment of the auxiliary
field allowed us also to define duality with respect to anomalous
subgroups by using the auxiliary field to cancel the anomaly.
This opens the way to perform duality with respect to independent
left and right groups which should have interesting applications
in heterotic $\sigma$-models. The general case without chiral
currents
is unfortunately still out of reach with our methods.
However we presented some moderately interesting examples of
non-abelian duality for four-dimensional gravitational
instantons. There are a number of open problems that we would
like to list here.
\begin{enumerate}
\item It could be interesting to work out the general structure
of the dual theory when there are no chiral currents. We are still
far from a satisfactory understanding of the general case.
\item We still do not have a clear picture of whether non-abelian
duality can be ``inverted'' as in the abelian case. Although in
the WZW or GKO-models we have characterized completely the dual
theory,
it is only a posteriori that we can deduce the starting theory.
\item The fact that we can explicitly characterize the non-abelian
dual of a large class of theories makes us suspect that there
should be an analog of this duality in lattice field theories.
However the explicit construction, and how the absence of
Pontrjagin duality for non-abelian groups is circumvented
remains to be explored.
\item The explicit operator mapping is still to be developed
fully. However we feel that the methods presented here
should help in completing this task.
\item The heterotic version of duality presented in section
five remains to be explored.
\item Dijkgraaf and Witten \cite{r24} showed in the context
of Chern-Simons theory how to gauge discrete subgroups in
the continuum. Given the close relation between Chern-Simons
theory and WZW-models, it seems reasonable to expect that a
combination of the methods presented in our paper together
with those in \cite{r24} should allow us to define duality
transformations with respect to discrete isometry groups.
If this is possible, a large and interesting class of examples
where one could try discrete duality is in Calabi-Yau
manifolds; of importance in String Theory
compactifications \cite{r25}.
\item The extension of our construction to supersymmetric
$\sigma$-models is also rather straightforward and it
will be presented elsewhere.
\end{enumerate}

\bigskip

{\large\bf Acknowledgements}

We would like to thank C. Burgess, E. Kiritsis, F.Quevedo,
G. Veneziano and E. Verlinde
for valuable discussions. E.A. and Y.L. were supported in part
by the CICYT grant AEN 93/673 (Spain) and by a fellowship of
Comunidad Aut\'onoma de Madrid (YL).

\bigskip

{\large\bf Note added}

After this work was completed we learned of the work \cite{r27}
where another attempt to understand non-abelian duality is made.
We also learned that in the note added to \cite{givroc} they
independently pointed out that the possible violation of
conformal invariance in the model of \cite{r7} should be related
to the fact that for non-semisimple groups there may be generators
in the adjoint representation whose trace is not zero. We would
like to thank M. Rocek for bringing this information to our
attention.

\end{document}